\documentclass[12pt]{article}
\usepackage{graphics,color}
\begin{document}
\thispagestyle{empty}
\noindent\
\\
\\
\\
\begin{center}
\large \bf  Composite Weak Bosons at the Large Hadron Collider
\end{center}
\hfill
 \vspace*{1cm}
\noindent
\begin{center}
{\bf Harald Fritzsch}\\
Department f\"ur Physik\\
Ludwig-Maximilians-Universit\"at\\
M\"unchen, Germany \\

\vspace*{0.5cm}
\end{center}

\begin{abstract}

In a composite model of the weak bosons the excited bosons, in particular the p-wave bosons, are studied. The state with the lowest mass is identified with the boson, which has been discovered recently at the "Large Hadron Collider" at CERN. Specific properties of the excited weak bosons are studied, in particular their decays into weak bosons and into photons.

\end{abstract}

\newpage

In the Standard Model of the electroweak interactions the masses of the weak bosons and of the leptons and quarks are generated by a spontaneous breaking of the electroweak symmetry. Besides the weak bosons a scalar boson must exist ( "Higgs boson"). Recently one has observed effects at the $LHC$, which might be due to the production of the Higgs boson and the subsequent decay into two weak bosons or into two photons (ref.(1,2)). The mass of this new boson is about 126 GeV.\\

In this lecture I discuss the possibility that the weak bosons are composite particles, i.e. particles with a finite size. The new scalar boson, observed at the $LHC$, would be an excited Z-boson, in which the constituents are in a p-wave.\\

In the Standard Theory of particle physics the leptons, quarks and the weak bosons are pointlike particles without any substructure. In the experiments at SLAC, DESY, Fermilab and CERN one has searched for a substructure of the electron, of the myon and of the light quarks, but nothing has been found. The present limit on the substructure of the electron, the muon and the quarks is about 0.001 fm.\\

Baryons and mesons are composed of quarks. Their extension is of the order of 1 fm. Their masses are directly linked to the substructure and can be calculated in the theory of quantum chromodynamics. The proton mass is due to the field energy of the gluons and quarks inside the proton, if the quark masses are set to zero. The field energy of the gluons provides about 70\% of the proton mass, 30\% are due to the field energy of the quarks. The mass of the proton is given by the scale parameter of QCD, multiplied by a dimensionless constant, which can be calculated:   $M(p)=const.\cdot\Lambda_{c}$.\\

In the Standard Theory the masses of the weak bosons, leptons and quarks are generated by the spontaneous symmetry breaking. A doublet of scalar fields is introduced, which breaks the weak isospin symmetry spontaneously and develops a non-zero vacuum expectation valus. The weak bosons absorb three of the four scalar fields and obtain a mass, which is proportional to the vacuum expectation value. The remaining neutral scalar boson is the "Higgs" boson, which might be the new boson, observed recently at the $LHC$.\\

Thus in the Standard Theory there are two different ways to generate masses. The masses of the hadrons are generated dynamically - the masses of the weak bosons and of the leptons and quarks originate from a spontaneous symmetry breaking. I prefer a theory, in which also the masses of the weak bosons and of the leptons and quarks are generated dynamically. This is possible, if the weak bosons as well as the leptons and quarks are composite particles.\\

About 60 years ago the $\rho$-mesons were considered to be pointlike gauge bosons, providing the strong interaction between the hadrons. The masses of the $\rho$-mesons can be introduced by a spontaneous symmetry breaking, analogous to the introduction of the masses of the weak bosons in the electroweak theory. A doublet of scalar fields is required, which describes four scalar particles. The three $\rho$-mesons absorb three scalars and become massive.\\

Such a mechanism for the mass generation of the $\rho$-mesons is now excluded. Their masses are dynamical masses as the nucleon mass. The $\rho$-mesons are not elementary gauge bosons, but quark-antiquark bound states. Besides the $\rho$-mesons there is also the iso-singlet $\omega$- meson, which could not obtain its mass through the spontaneous symmetry breaking, involving a scalar doublet.\\

In QCD the three $\rho$-mesons are degenerate in mass, if the electromagnetic interaction is switched off and the two light quark masses are zero. Once the electromagnetic interaction is introduced, the charged mesons receive an additional small contribution to the mass, which is due to the Coulomb self energy. In addition the neutral $\rho$-meson mixes with the photon and its mass increases. This mass shift can be calculated. It depends on a mixing parameter $\mu$, which is determined by the electric charge, the decay constant $F_\rho$ and the mass of the $\rho$-meson:\\

\begin{equation}
\mu \; =\; e \frac{F^{}_\rho}{M^{}_\rho}~ .
\end{equation}\\

The mass shift due to the mixing is given by:\\

\begin{equation}
M^2_{\rho^0} \; =\; M^2_{\rho^+} \left(\frac1{1 - \mu^2}\right) \; .
\end{equation}\\

The decay constant is measured to about 220 MeV - it is about equal to the  QCD scale parameter $\Lambda_{c}$. One obtains $\mu\approx 0.09$ - it leads to a mass shift of about 3 MeV. Due to this mass shift the mass difference between the charged and neutral $\rho$-meson is rather small.\\

We assume that the weak bosons are composite particles. They consist of a lefthanded fermion and its antiparticle, which are denoted as "haplons".  A theory of this type was proposed in 1981 (see  ref.(3) and ref.(4,5,6,7,8)). The new confining chiral gauge theory is denoted as $QHD$. The $QHD$ mass scale is given by a mass parameter $\Lambda_h$, which determines the size of the weak bosons. The haplons interact with each other through the exchange of massless gauge bosons, which we denote as "glutons" (the Latin expression for "glue" is "gluten").\\

Two types of haplons are needed as constituents of the weak bosons, denoted by $\alpha$ and $\beta$.
Their electric charges in units of e are:\\

\begin{equation}
h = \left( \begin{array}{l}
+\frac{1}{2}\\
-\frac{1}{2}\\
\end{array} \right) \ .
\end{equation}\\

The three weak bosons have the following internal structure:\\
\begin{eqnarray}
W^+ & = & \overline{\beta} \alpha \; , \nonumber \\
W^- & = & \overline{\alpha} \beta \; , \nonumber \\
W^3 & = & \frac{1}{\sqrt{2}} \left( \overline{\alpha} \alpha -
\overline{\beta} \beta \right) \; .
\end{eqnarray}\\
In the absence of electromagnetism the weak bosons are degenerate in mass. If the electromagnetic interaction is introduced, the mass of the neutral boson increases due to the mixing with the photon (ref. (9,10)).\\

In the Standard Theory the mixing is generated by the scalar fields. Both the photon and the Z-boson are mixtures of the $SU(2)$ and $U(1)$ gauge bosons. Here the mixing is a dynamical mixing, analogous to the mixing of $\rho$ - mesons. It is described by the mixing parameter m, determined by the decay constant of the weak boson:

\begin{equation}
m =\; e \frac{F^{}_W}{M^{}_W}~ .
\end{equation}\\

One finds for the mass difference beween the charged and the neutral weak boson:\\

\begin{equation}
M^2_{Z} \; =\; M^2_{W^+} \left(\frac1{1 - m^2}\right) \; .
\end{equation}\\

In the standard electroweak theory there is a similar equation - the mixing parameter m must be replaced by $\sin\theta_w$. According to the experiments the mixing parameter m is about 0.485, i. e. about five times larger than the mixing paramter for the $\rho$-mesons. Using the experimental value, one can determine the decay constant for the weak bosons:

\begin{equation}
F^{}_W\approx125\,GeV.
\end{equation}\\

As in $QCD$ it is expected that the decay constant of the weak boson and the $QHD$ mass scale are related. The decay constant of the $\rho$-meson and the $QCD$ mass scale are about the same - in $QHD$ the weak decay constant and $\Lambda_h$ should be of the same order of magnitude. Details will depend in particular on the gauge group of $QHD$. We expect that $\Lambda_h$ is in the range between 0.12 TeV and 1 TeV.\\

At low energies our composite model and the standard elektroweak gauge theory are essentially identical. But at high energies large deviations are expected. In particular a new isoscalar neutral weak boson $X$ should exist, which is analogous to the $\omega$-meson:
\begin{eqnarray}
X & = & \frac{1}{\sqrt{2}} \left( \overline{\alpha} \alpha +
\overline{\beta} \beta \right) \; .
\end{eqnarray}\\
The present limit on the mass of this isoscalar weak boson is about 0.4 TeV. It will decay mainly into weak bosons, e.g. into two charged weak bosons or into two $Z$-bosons. Decays of the $X$-boson into two fermions are expected to be strongly suppressed. We expect that the mass of the $X$-boson is less than 1 TeV, thus it should be observed soon at the $LHC$.\\

 Above the energy of 1 GeV exist many excited states of the baryons and mesons. We expect similar effects in the electroweak sector. The  $QHD$ mass scale is about thousand times larger than the  $QCD$ mass scale. Thus at energies above 0.1 TeV there should in particular exist excited weak bosons, which will decay mainly into several weak bosons.\\

The weak bosons consist of pairs of haplons, which are in an s-wave. The spins of the two haplons are aligned, as the spins of the quarks in a $\rho$-meson. The first excited states are those, in which the two haplons are in a p-wave. We describe the quantum numbers of these states by $I(J)$. The $SU(2)$-representation is denoted by $I$ - the symbol $J$ describes the total angular momentum. There are three $SU(2)$ singlets, which we denote by S(0), S(1) and S(2):\\
\begin{eqnarray}
S(0)=[0(0)], \nonumber \\
S(1)=[0(1)], \nonumber \\
S(2)=[0(2)],
\end{eqnarray}\\
\\
and three triplet states:\\
\begin{eqnarray}
T(0)=[1(0)], \nonumber \\
T(1)=[1(1)], \nonumber \\
T(2)=[1(2)].
\end{eqnarray}\\
\\
These three states have the internal structure:\\
\begin{eqnarray}
T^+ & = & \overline{\beta} \alpha \; , \nonumber \\
T^- & = & \overline{\alpha} \beta \; , \nonumber \\
T^3 & = & \frac{1}{\sqrt{2}} \left( \overline{\alpha} \alpha -
\overline{\beta} \beta \right) \; .
\end{eqnarray}\\
We compare the spectrum of these excited bosons with the spectrum of the mesons in strong interaction physics, in which the quarks are in a p-wave. The lowest meson is the scalar $\sigma$-meson (mass  $ \sim 600  -  800 $ MeV). This meson can easily mix with glue mesons, i.e. the $\sigma$-meson would be a superposition of a quark-antiquark-meson and a glue meson. This might explain the relatively low mass of this meson.\\

The $\sigma$-meson is the QCD analogue of the boson $S(0)$. The spin one meson with isospin zero is $h_1(1170)$, analogous to the excited boson S(1). The corresponding spin two meson is $f_2(1270)$. It is analogous to the excited boson S(2):\\

\begin{eqnarray}
 S(0) \sim \sigma, \nonumber \\
 S(1) \sim h_1, \nonumber \\
 S(2) \sim f_2.
\end{eqnarray}\\
We expect that the scalar state $S(0)$ with the quantum numbers $[0(0^{+})]$ mixes with gluton bosons, analogous to the mixing of the $\sigma$-meson with glue mesons. Due to this mixing the mass of $S(0)$ is much lower than the masses of the other p-wave states.\\

Now we consider the isospin triplet mesons, the scalar meson $a_0(980)$, the vector meson $b_1(1235)$ and the spin two meson $a_2(1320)$. These mesons are analogous to the bosons T(0), T(1) and T(2) respectively:\\

\begin{eqnarray}
 T(0) \sim a_0, \nonumber \\
 T(1) \sim b_1, \nonumber \\
 T(2) \sim a_2.
\end{eqnarray}\\
\\
What is the mass spectrum of the excited weak bosons, in particular of the $S$ and $T$ bosons? The boson $S(0)$ should have a relatively small mass, due to the mixing with the gluton bosons. We identify this state with the particle, which has been observed at CERN (ref. (1,2)): \\

\begin{equation}
M(S(0))= 126~GeV.
\end{equation}\\

In analogy to QCD we expect that the masses of the other p-wave states are in the range 0.3 - 0.5 TeV. The mixing of the bosons $S(1)$ and $S(2)$ with the gluton bosons is not as strong as the mixing for the $S(0)$. Thus we expect that the mass of the $S(1)$ - boson is just above 0.3 TeV, and the mass of the $S(2)$ - boson would be between 0.4 and 0.5 TeV.\\

The $SU(2)$ - triplet bosons $T$ cannot mix with gluton bosons. For this reason their masses should be larger than the masses of the $S$ - bosons. We compare the spectrum of these bosons with the spectrum of the corresonding mesons in $QCD$. Thus the mass of the $T(0)$ - boson should be about 0.3 TeV, the mass of the $T(1)$ - boson just above 0.4 TeV, and the mass of the  $T(0)$ - boson should be in the range 0.5  - 0.6 TeV.\\

The $S(0)$ - boson will decay mainly into two charged weak bosons or into two $Z$-bosons (one of them virtual respectively), into a photon and a $Z$-boson and into two photons. The $Z$-boson is the boson $W^3$, mixed with the photon. The mixing angle is the weak angle, measured to about 28.7 degrees. Using this angle, we can calculate the branching ratios BR for the various decays - the branching ratio for the decay into charged weak bosons is denoted by B.\\

 First we assume that the $S(0)$ - boson has a mass, which is more than twice the mass of the $W$ - boson, in order to avoid phase space calculations. We find for the branching ratios:\\

$S(0) \Longrightarrow  (W^+ + W^-) $    ~~~~~~~~~                BR = $B$,\\

$S(0) \Longrightarrow  (W^- + W^+) $     ~~~~~~~~~              BR = $B$,\\

$S(0) \Longrightarrow  (Z + Z) $     ~~~~~~~~~              BR  $\simeq 0.59 ~B $, \\

$S(0) \Longrightarrow  ( Z +  \gamma ) $      ~~~~~~~~~~~~~~~~~~             BR $\simeq 0.18 ~B $, \\

$S(0) \Longrightarrow  (\gamma + \gamma) $    ~~~~~~~~~~~~~~~~~~~      Br $\simeq 0.05 ~B$.\\

Now we consider the real value of the mass of the  $S(0)$ - boson (126 GeV), which implies that one of the weak bosons can only be produced as a virtual particle, decaying into a fermion pair (a virtual weak boson is indicated by "W" or "Z"). Taking into account phase space corrections, one finds: \\

$S(0) \Longrightarrow  ("W^+" + W^-) $    ~~~~~~~~~               BR = $B$,\\

$S(0) \Longrightarrow  ("W^-" + W^+) $     ~~~~~~~~~              BR = $B$,\\

$S(0) \Longrightarrow  ("Z" + Z) $     ~~~~~~~~~              BR  $\approx 0.35 ~B $, \\

$S(0) \Longrightarrow  ( Z +  \gamma ) $      ~~~~~~~~~~~~~~~~~~             BR $\approx 0.10 ~B $, \\

$S(0) \Longrightarrow  (\gamma + \gamma) $    ~~~~~~~~~~~~~~~~~~~        Br $\approx 0.06 ~B$. \\

We emphasize that the branching ratios for the decays into two $Z$ -bosons, into two photons and into a photon and a $Z$ - boson have rather large errors, due to the uncertainties in the phase space calculations.\\

We would not expect that the decay rates for the decays of $S(0)$ into letons and quarks are given by the mass of the fermion, as they are for the Higgs boson. The branching ratio for the decay into an electron pair, into a muon pair, into a tau pair or into a neutrino pair would be similar: \\

BR ($S(0) \Longrightarrow  (e^+ + e^-) $ ) $\approx $ BR ($S(0) \Longrightarrow  (\mu ^+ + \mu^-) $ ) \\

$\approx $ BR ($S(0) \Longrightarrow  (\tau ^+ + \tau^-) $ ) $\approx $ BR ($S(0) \Longrightarrow  (\nu + \bar{\nu}) $ ), \\

( $\nu$ stands for the electron, muon and tau neutrino ). \\

Due to the color of the quarks the branching ratio for the decays into quarks are three times larger:\\

BR ($S(0) \Longrightarrow  (u + \bar{u}) $ )  $\approx $ 3 BR ($S(0) \Longrightarrow  (e^+ + e^-) $ ) \\

BR ($S(0) \Longrightarrow  (d + \bar{d}) $ )  $\approx $ 3 BR ($S(0) \Longrightarrow  (e^+ + e^-) $ )\\

BR ($S(0) \Longrightarrow  (u + \bar{u}) $ ) $\approx $ BR ($S(0) \Longrightarrow  (c + \bar{c}) $ )\\

BR ($S(0) \Longrightarrow  (d + \bar{d}) $ ) $\approx $ BR ($S(0) \Longrightarrow  (s + \bar{s}) $ ) \\

$\approx $ BR ($S(0) \Longrightarrow  (b + \bar{b}) $ ).\\

Presumably the $S(0)$ decays mainly into two weak bosons. But decays of the $S(0)$ into muon pairs could be observed in the near future at the $LHC$.\\

The bosons $S(1)$ and $S(2)$ have a much higher mass as the $S(0)$ - boson. They will decay mainly into three or four weak bosons, e.g.: \\

$S(1) \Longrightarrow  (W^+ + W^- + Z) $,\\
$S(1) \Longrightarrow  (W^+ + W^- + W^+ + W^-) $,\\
$S(1) \Longrightarrow  (W^+ + W^- + \gamma) $.\\

Decays of $S(1)$ or $S(2)$ into two weak bosons or two photons would be suppressed.\\

The $SU(2)$ - triplet bosons $T(0)$, $T(1)$ and $T(2)$ will decay mainly into four or five weak bosons or photons. Decays into two weak bosons, a weak boson and a photon or two photons are strongly suppressed.\\

The boson $T^+$ or $T^-$ can in principle decay into a charged weak boson and a $Z$-boson or a photon, however the branching ratio for this decay will be very small. The main decay modes of the $T^+$ would be:\\

$T^+ \Longrightarrow  (W^+ + Z + Z) $,\\

$T^+ \Longrightarrow  (W^+ + Z + \gamma ) $,\\

$T^+ \Longrightarrow  (W^+ + \gamma + \gamma) $.\\

The properties of the new boson with a mass of 126 GeV, which has been discovered at the $LHC$  ,  should be investigated in detail. If the model, discussed here, is correct and the new boson is the state $S(0)$, the other excited bosons $S(1)$, $S(2)$ and $T(0)$ should be discovered in the near future at the LHC.\\
\\
\\
\\

\end{document}